\documentclass[MSNbibl,nameyear,dvips]{arxstspdf}
\usepackage{flushend}
\usepackage{stfloats}
\volume{29}
\issue{1}
\pubyear{2014}
\firstpage{104}
\lastpage{105}
\doi{10.1214/13-STS463} % kopijuoti is 'New paper accepted'
\referstodoi{10.1214/13-STS420}% pagrindinio straipsnio DOI, kai
\begin{document}
\begin{frontmatter}
\vspace*{6pt}
\title{Response to Discussion by A. H. Welsh on the AF 447 Paper}%
% kai straipsnis turi susijusiu diskusiju ir rejoinder'iu
%rejoinder at \relateddoi{r}{10.1214/00-STSXXXX}.}
\runtitle{Response to A. H. Welsh}

\begin{aug}
\author[a]{\fnms{Lawrence D.} \snm{Stone}\corref{}\ead
[label=e1]{stone@metsci.com}\ead[label=u1,url]{www.metsci.com}}
\runauthor{L. D. Stone}

\affiliation{Metron, Inc.}

\address[a]{Lawrence D. Stone is Chief Scientist, Metron, Inc., 1818
Library Street, Suite 600, Reston, Virginia 20190, USA
\printead{e1,u1}.}

\end{aug}

% ABSTRACT

% KEYWORDS
% Pirmas kwd is didziosios raides

\end{frontmatter}

I thank Professor Welsh for his very kind comments about the AF 447
paper. He makes a number of excellent points. One is that Bayesian
analysis is a tool and that must be used carefully and thoughtfully in
order to obtain good results in a complicated problem such as the
search for AF 447. While this is true, the use of Bayesian analysis is
required to incorporate the necessary subjective judgments into the
analysis of the AF 447 search. As Welsh notes, Bayesian analysis
allowed us to propagate these judgments and uncertainty distributions
into the probability distribution on the location of the wreck in a
logical and correct fashion. Classical statistics does not provide a
framework for doing this. Bayesians should celebrate this advantage.

The power of Bayesian analysis as a tool is further illustrated by the
U.S. Coast Guard's Search and Rescue Optimal Planning System (SAROPS).
SAROPS is a Bayesian search planning program used by the Coast Guard
every day for planning searches for people and boats lost at sea. It is
run by Coast Guard officers who are trained to use the program but are
by no means experts in Bayesian analysis. The Coast Guard considers it
one of their best operational computer programs.

Welsh suggests that the use of data from nine somewhat similar
situations casts doubt on the claim that the use of subjective
probabilities is required for the AF 447 analysis. However, the
availability of this data does not mean we could reasonably have
produced the AF 447 distribution without the use of subjective
probabilities. The use of subjective probabilities is one
characteristic that distinguishes Bayesian statistics from classical
statistics where decisions are supposed to be made solely on the basis
of objective information and scientific analysis. It seems to me that
Bayesian analysis is uniquely suited for tackling complicated problems
of this sort.

Welsh asks two interesting questions: (1) What would be the result of a
Bayesian version of the reverse drift analysis (performed by the drift
group) that produced the rectangle for the fourth unsuccessful search?
(2) What is the correct way to handle the uncertainty about whether the
underwater locator beacons functioned or not?

Question (1) is answered in the paper. The process of producing the
reverse drift scenario distribution was our attempt to do the reverse
drift analysis in a Bayesian fashion accounting for the uncertainties
in the winds, currents and drift behavior of dead bodies. This analysis
produced a distribution that spread over a very large area of the
ocean. When we intersected this distribution with the 40 NM circle, we
obtained the distribution shown in Figure~3 of the paper. In
retrospect, it appears that this would have been a pretty good prior
distribution for the location of the wreck before any search took
place. By comparison, the rectangle produced by the drift group is in a
very low probability region of this distribution. The ``uncertainties
in the uncertainties'' in the reverse drift scenario distribution would
have given us pause in recommending it as the sole method of computing
the prior location distribution. In computing this distribution, we
used the drift group's choice for the best current estimate, but there
were other possibilities that were reasonable too. The estimate
provided only a mean current without any stochastic component to it. We
had to add uncertainties to the mean in order to obtain a stochastic
process for the currents. These uncertainties coupled with the large
spread in the resulting location distribution left us with low
confidence in this scenario.

Question (2) is also answered in the paper. At the end of Section~4.6,
we note that ``a better way to handle the doubts we had about the
beacons would have been to compute a joint distribution on beacon
failure and wreck location. The marginal\vadjust{\goodbreak} distribution on wreck location
would then be the appropriate posterior on which to base further
search.'' After the unsuccessful passive search, the joint posterior
distribution would have reflected correctly both the possibility that
beacons were not working and that they were working but not
detected.
The marginal distribution on beacon failure would have provided a
quantitative estimate of the probability of beacon failure. Providing
the joint distribution would\vadjust{\eject} have been better than providing the BEA
with two distributions, one assuming the beacons functioned and one
assuming they failed.

The passive search did indeed cover the location of the wreck. If the
beacons\ had been working properly, it is highly likely that the passive
acoustic search would have detected them and that Bayesian analysis and
the authors of the AF 447 paper would never have been involved in the search.

% zodis "Acknowledgments" paliekamas pagal autoriu

%suskaldyti doi

\end{document}